\documentstyle[aps,prl,multicol,epsfig]{revtex}
\begin{document}
\title {Inversion of Optical Reflectance in the Fullerenes}
\def\bea {\begin{eqnarray}}
\def\eea {\end{eqnarray}}
\def\be {\begin{equation}}
\def\ee {\end{equation}}
\def\eqnum#1{\eqno (#1)}
\def\sup{superconductivity }
\def\high{high $T_c$ superconductivity }
\author{F. Marsiglio}
\address {Department of Physics, University of Alberta, Edmonton, AB 
T6G 2J1}

\maketitle
\begin{abstract}
Since the discovery of superconductivity in the alkali-doped fullerenes
\cite{hebard}, the electron-phonon interaction has been the primary
suspect for superconductivity in this class of compounds. In this paper 
we first provide a pedagogical review of how the question
of mechanism has traditionally been settled, 
and then some well-known properties of the optical reflectance
(and the derived conductance) are summarized. 
Finally we demonstrate how a recently derived
inversion procedure can use the optical properties of a metal to infer the
magnitude of the electron-phonon interaction. We conclude that in the case
of the alkali-doped fullerenes, the electron-phonon interaction is sufficiently
strong to explain the transition temperatues found in these the materials.
\end{abstract}\par
\bigskip
\begin{multicols}{2}
\bigskip
\noindent {\it Introduction}
\medskip
\par
The fcc solid composed of C$_{60}$ molecules at the lattice sites is an
insulator, and almost certainly a band insulator \cite{satpathy}.
As is the case with graphite, this material can be doped with alkali metal 
atoms. One can try for various
`stoichiometric' results, such as A$_n$C$_{60}$, where A is an alkali metal
atom, and $n$ is an integer ranging from 1 through 6. Various interesting phases
result, as a function of $n$ (and to a lesser extent A). In some cases, the
compound apparently doesn't form, with indications of phase separation at
low temperature (see Ref. \cite{dresselhaus}, for example). The point of interest
for us is $n = 3$, for it is at this doping level (and apparently only this
doping level) that superconductivity occurs. Unlike the focus of much of the
work in high temperature superconducting cuprate materials (which also have an
insulating parent state), it appears to be a much more difficult problem here
to relate the various phases that occur as a function of $n$. Here the situation
is intrinsically different, in that only stoichiometric compounds form (although
suggestions that this may also be the case in the cuprates have appeared from 
time to time). Thus, for this work we relinquish the `big picture', and focus
only on A$_3$C$_{60}$, and in particular, A = K, which is a superconductor with
a transition temperature, $T_c = 20 $ K.\par

There are many indications \cite{dresselhaus,reviews} that the K$_3$C$_{60}$
compounds are (a) well-behaved Fermi Liquids, and (b) conventional 
electron-phonon superconductors. Property (a) does not imply property (b),
although property (b) presupposes that (a) is true.
The main indication that Fermi Liquid theory (FLT) might be the appropriate
starting point is the temperature dependence of the resistivity, which
exhibits the standard FLT behaviour, $ \rho (T) = a + bT^2$ \cite{xiang}. 
Various measurements below the critical temperature indicate that the
superconducting state is a conventional s-wave one. Gap measurements, using
both tunneling and infrared spectroscopy \cite{zhang,koller,degiorgi} indicate
an s-wave gap, though there is some controversy over the size of the gap.
The existence of an s-wave gap is also supported by penetration depth
measurements \cite{uemura}. Furthermore, a coherence peak in the spin 
relaxation rate, $1/T_1$, has been measured using muons \cite{keifl}, 
a result which lends strong support to an s-wave BCS picture.
Finally, the isotope effect has been determined to be significant 
(see Ref. \cite{franck} for a summary), although again the size of the
isotope exponent is controversial.  

\bigskip
\noindent {\it The Question of Mechanism}
\medskip
\par
All of these measurements support the existence of an s-wave gap, but,
with the exception of the isotope effect, have little to say about
mechanism. Partly this is the ``curse'' of universality; BCS theory was
so successful partly because it provided universal results (for various
properties and temperature dependences). However, one consequence of this
is that for a wide variety of mechanisms, if the coupling is weak, the
prediction for many superconducting properties is the same. Hence, the
theory is not very useful for discriminating mechanism. One can claim,
for example, that in the case of Al (which is a very weak coupling
superconductor), there is no definite proof for electron-phonon
superconductivity \cite{rainer88}. For certain
superconducting materials (known in the older literature as ``bad actors'')
like Pb and Hg, deviations from universal behaviour were measured, so there
was hope that these deviations could point towards a mechanism. This is
exactly what happened in the case of Pb, particularly through tunneling
experiments, as we will now describe. 
 
In single particle tunneling experiments an (insulating) oxide layer is
sandwiched between a superconducting material and a normal metal (other
variants are possible, but we focus on this, the simplest). An external circuit
allows electrons to tunnel across this barrier. If a single particle energy
gap exists in the superconducting side, then no current will flow until
the bias voltage exceeds this energy gap. In this way one can measure the
energy gap in a superconductor (provided one exists). This is exactly what
the original experiments by Giaever \cite{giaever} measured (in an Al/Pb
sandwich).

Actually, even this measurement indicated that something was wrong. The
gap ratio, $2\Delta/k_BT_c$, was measured to be well over 4 (later
more accurate measurements indicated a value closer to 4.5), whereas
BCS theory predicts a universal result, 3.5. Strong coupling was suspected,
although strong coupling extensions of BCS theory only managed to raise
this universal value to yet another universal maximum value, 4.0 \cite{swihart}.
By this time Eliashberg had extended the theory to include dynamical
interactions with phonons \cite{eliashberg}. As further refinements in
both theory and experiment developed, it became clear that certain structure
in the density of states (as measured by electron tunneling in Pb, one of the
``bad actors'') was due to the electron-phonon interaction 
\cite{rowell63,schrieffer63}. Finally, McMillan and Rowell succeeded in
using tunneling measurements and the Eliashberg equations to extract the
underlying electron-phonon interaction in the form of the electron-phonon
spectral function, $\alpha^2F(\Omega)$ \cite{mcmillan65,mcmillan69}.   
We now describe qualitatively the essence of Eliashberg theory (for various
reviews see Ref. \cite{scalapino69,allenmitrovic,rainer86,carbotte90}), and of
the McMillan-Rowell inversion procedure.

Within BCS theory, the gap parameter $\Delta_{\hbox{BCS}}$ serves both as 
an order parameter, and as the energy below which the density of 
states is zero (hence the name ``gap'' parameter). The only input
required in the theory is a model attractive potential, multiplied by
the electron density of states at the Fermi level: $N(\epsilon_F)V$. As
already mentioned above, Eliashberg theory goes one step further; it
recognizes that a source for the attraction is the electron-phonon
interaction, and that retardation effects can be important since the
mean ion velocity is significantly less than the Fermi velocity of the
electrons. The result is that the source of attraction is encapsulated by
an electron-phonon spectral function, 
$\alpha^2F(\Omega)$. Moreover, the order parameter in Eliashberg theory is 
frequency dependent. Finally, the direct electron-electron Coulomb repulsion
is recognized as a source of depairing, and is generally included at a
minimal level through a single dimensionless parameter, $\mu^\ast$. 
The `$\ast$' is present to remind us that this is a pseudopotential, 
and ought to have a greatly reduced value if one uses cutoffs in the
theory which apply to the phonon energy scale. To summarize, Eliashberg
theory produces an equation analagous to the BCS one, except that the
order parameter is a function of frequency $\omega$, a functional of 
$\alpha^2F(\Omega)$, and a function of $\mu^\ast$. It is also complex
(not in the usual sense: this applies even if the phase has been set to zero).
Once this gap function, $\Delta (\omega)$, is obtained, one can
calculate a variety of properties, one of which is the single particle
density of states, $N(\omega) = N(\epsilon_F) \hbox{Re} \bigg\{ {\omega
\over \sqrt{\omega^2 - \Delta^2(\omega)} }  \bigg\}$. This expression
bears a remarkable resemblance to the weak coupling result, for which the
gap function would simply be replaced by the gap parameter, $\Delta_{\hbox{BCS}}$. 
Note that the I-V characteristic is proportional to the single particle
density of states: $dI(V)/dV \propto N(V)$, so that an I-V measurement immediately
yields ths single particle density of states.

The procedure formulated by McMillan and Rowell is the following. One first
measures the structure above the gap edge in $dI/dV$ accurately. Next,
one ``guesses'' an $\alpha^2F(\Omega)$, and using this guess computes
$\Delta (\omega)$, and hence $dI(\omega)/dV$, using Eliashberg theory.
The calculated function is compared to the experimental result, and
corrections to the initial guess are computed. Corrections can be obtained
through functional derivatives; the procedure amounts to a Newton-Raphson
method for functions (rather than for single variables). The new
$\alpha^2F(\Omega)$ is used through the procedure all over again, until
the result converges. One then claims to have ``measured'' the 
underlying $\alpha^2F(\Omega)$.
That this procedure works is a priori not obvious. There seems to be no reason
why the solution should be unique, but apparently most of the time it is.
That the result is physically meaningful is apparent because of several
tests. First, that the function comes out positive is already a triumph.
Negative spectral functions are not ruled out in the iterative 
process, so the positivity indicates a physical
result (the spectral function must be positive (for positive frequency)).
Although not apparent from the above description, the inversion procedure
requires knowledge of the density of states in the phonon region only.
Nonetheless, structure occurs well beyond the phonon region. One can then use
the spectral function with Eliashberg theory to {\it predict} the density
of states at higher energies. The agreement with the experimental result is
spectacular (see Fig. 32 in Ref. \cite{mcmillan69}). Finally one can compare
to neutron scattering experiments, which measure the phonon density of states,
$F(\Omega)$, directly; these experiments tell us at what energies the 
phonons exist, but not how strongly coupled they are to the electrons. Hence,
the two functions, $F(\Omega)$, and $\alpha^2F(\Omega)$, don't have to
be similar in many respects, though the frequency range of the latter clearly
cannot exceed that of the former. As it turns out, these functions are rather
similar (and, of course, fulfill the required condition); this indicates
that the coupling function $\alpha^2(\Omega) \equiv \alpha^2F(\Omega)/F(\Omega)$,
is not very frequency dependent. A detailed comparison of these two probes
is provided in Ref. \cite{rowell71}. 

Finally we should mention that other measurements often corroborate this
story. The gap ratio often comes out in better agreement with experiment,
and various thermodynamic quantities show better agreement too \cite{carbotte90}.
One might well wonder at how successful this approach has been historically,
though, since important effects apparently have been omitted. For example,
electron-electron Coulomb interactions have only been included insofar as they
alter the underlying band structure (this is implicit), and through the
single parameter $\mu^\ast$ (this is explicit in the theory).

Perhaps partially for the reasons mentioned in the previous paragraph, along
with other aspects, such tunneling measurements have so far failed to
provide conclusive results for the cuprate superconductors. In the case of
the alkali-doped fullerenes, tunneling has clearly demonstrated a gap,
but so far tunneling measurements have not been used to invert for
$\alpha^2F(\Omega)$ \cite{ostrick}. Part of the problem is that the phonons
are known from neutron scattering \cite{pintschovius} to extend out to 
about 250 meV, much further than in conventional superconductors. 
For these reasons we looked for an alternate method to deduce the underlying
interactions, one which uses the measured optical conductivity \cite{marsiglio98}. 
In the next section we briefly describe optical measurements and then describe
our procedure and its application to K$_3$C$_{60}$.

\bigskip
\noindent {\it Optical Properties in Metals}
\medskip
\par
A common method to determine the optical conductivity is to measure the 
reflectance \cite{timusk} as a function of frequency, usually at normal 
incidence. The reflectance, $R(\nu)$, is defined as the absolute ratio 
squared of reflected over incident electromagnetic wave amplitude. 
The complex reflectivity is then defined by
\be
r(\nu) \equiv R(\nu)\exp{(i\theta(\nu))},
\label{reflectivity}
\ee
where $\theta(\nu)$ is the phase, and is obtained through a Kramers-Kronig
relation from the reflectance \cite{timusk}
\be
\theta(\nu) = {\nu \over \pi} \int_0^\infty {\ln{R(\nu^\prime)} - \ln{R(\nu)}
\over \nu^2 - {\nu^\prime}^2 } d\nu^\prime.
\label{kk}
\ee
The complex reflectivity can be related to the complex index of refraction,
$n(\nu)$,
\be
r(\nu) \equiv {1 - n(\nu) \over 1 + n(\nu)},
\label{index}
\ee
which, finally, is related to the complex conductivity, $\sigma(\nu)$ (using
the dielectric function, $\epsilon(\nu)$:
\be
\epsilon(\nu) \equiv n^2(\nu) = \epsilon_\infty + {4 \pi i \sigma(\nu) \over
\nu},
\label{dielectric}
\ee
where $\epsilon_\infty$ is the dielectric function at high frequency (in principle,
for infinite frequency this would be unity). It is through such transformations
that the `data' is often presented in `raw' form. Nonetheless, assumptions are
required to proceed through these steps; for example, Eq. (\ref{kk}) indicates quite
clearly that the reflectance is required over all positive frequencies. Thus
extrapolation procedures are required at low and high frequencies; a more
thorough discussion can be found in \cite{wooten}; see also \cite{marsiglio97}.

In Figs. 1 and 2 we illustrate how elastic and inelastic scattering processes
affect the reflectance. In Fig. 1 we use a simple Drude form for the complex
conductivity, $\sigma(\nu) = \sigma_0/(1 - \nu \tau)$ and plot the reflectance 
which results from the application of the above formulae. Here the coefficient
$\sigma_0$ is the dc conductivity, and the parameter $1/\tau$ is the electron 
elastic scattering rate. The source of elastic scattering is
unspecified, but usually results from electron-impurity scattering. In Fig. 1
the plasma edge is evident, and occurs, of course, near the plasma frequency,
$\omega_P$. The "sharpness" of the dropoff decreases with
\narrowtext
\begin{figure}
\mbox{
\epsfig{figure=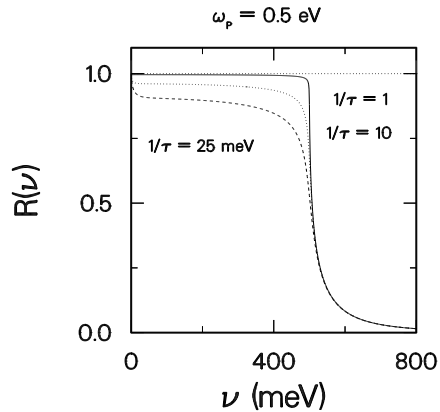,width=0.4\textwidth,clip=}}
\caption{The reflectance $R(\nu)$ vs. frequency for the Drude model.
Three curves are shown, corresponding to the three elastic scattering
rates shown.}
\label{fig1}
\end{figure}

\begin{figure}
\mbox{
\epsfig{figure=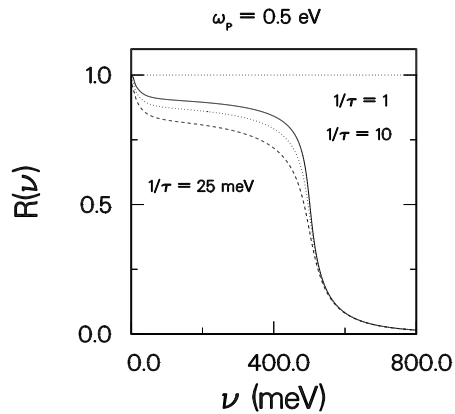,width=0.4\textwidth,clip=}}
\caption{The reflectance $R(\nu)$ vs. frequency for the case shown in
Fig. 1, but with additional inelastic scattering (modelled through
the electron-phonon interaction in Pb). Note the general rounding of
the curves and the decreased reflectance (as compared with those in Fig. 1).}
\label{fig2}
\end{figure}

\noindent increasing scattering
rate. The plateau region below the plasma frequency is referred to as the
relaxation region; the level decreases with increasing scattering rate, as
expected, since stronger scattering should cause the reflectance to decrease.
Finally, at very low frequency, the Hagen-Rubens form, $R(\nu) \approx
1 - \sqrt{{8 \over \omega_P \tau}{\nu \over \omega_P}}$ takes the reflectance
smoothly to unity at zero frequency.

In Fig. 2 we use the same elastic scattering rates for the three curves
shown in Fig. 1. In addition we include inelastic scattering through the
electron-phonon interaction (in this case with Pb). Note that the main
effect is a rounding of the edge, and a general decrease of the reflectance
everywhere. The effect is very nonlocal in frequency --- note that the
Pb phonon spectrum extends to about 10 meV, which is tiny on the scale of
this figure, and particularly on the frequency scale of the changes with
respect to Fig. 1. The effect of the inelastic scattering is also
cumulative, a result which follows from the fact that the effective scattering
rate increases as a function of frequency \cite{dolgov91,marsiglio97}.

The corresponding conductivity curves are shown in Figs 3-5. In Fig. 3
we show the Drude form (Lorentzian) for the three scattering rates shown
previously. The results shown in Fig. 3 are independent of temperature.
In Fig. 4 we suppress the elastic scattering rate ($1/\tau = 0$), to
see more clearly the effect of inelastic scattering (here included through
the electron-phonon interaction in {Ba$_{1-x}$K$_x$BiO$_3$}, with $T_c = 29$ K).
The results shown are in the normal state only; we have referenced the
temperatures to the critical temperature as a convenience only. What is
clear is that at high temperatures, enough phonons are present to mimic 
impurity scattering, so that a ``Drude-like'' peak is present. At low
temperatures this peak vanishes, and structure becomes more visible (solid
curve in the figure). This structure provides a signature of the electron-phonon
interaction, analogous to the information inherent in the structure in the
tunneling I-V characteristic \cite{marsiglio95,dolgov95}. 
Finally, in Fig. 5, we show similar results, but with an elastic scattering
rate $1/\tau = 10$ meV, so that a Drude peak remains even at zero
temperature. Note that the structure remains, but is not nearly as
visible as in the clean limit, and neither seems to display the degree
of structure seen in tunneling. We also remark that similar statements apply to
the imaginary part of the conductivity (see Ref. \cite{marsiglio96} for
a full discussion).

\bigskip
\noindent {\it The Inversion of the Optical Conductivity}
\medskip
\par

The inference of the electron-phonon interaction through the optical
conductivity has a long history. No doubt it was apparent that the
information was available right from the initial studies of Holstein
\cite{holstein54}, where he made clear that inclusion of the
electron-phonon interaction would have a significant impact on the
optical conductivity. The first real attempt, however, to infer

\begin{figure}
\mbox{
\epsfig{figure=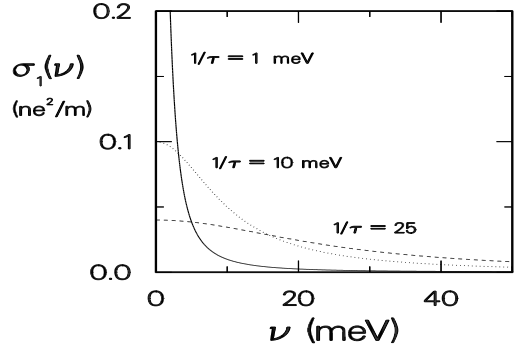,height=8.5cm,width=0.5\textwidth,clip=}}
\caption{The real part of the optical conductivity in the Drude model (elastic
scattering only) for the same three scattering rates shown in Fig. 1.
These results are independent of temperature.}
\label{fig3}
\end{figure}

\begin{figure}
\mbox{
\epsfig{figure=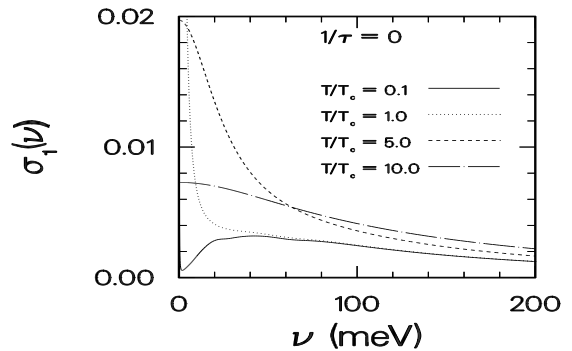,height=8.5cm,width=0.5\textwidth,clip=}}
\caption{The real part of the optical conductivity in the clean limit (no
elastic scattering). The only source of scattering is inelastic, as modelled by
the  electron-phonon interaction in Ba$_{1-x}$K$_x$BiO$_3$ ($T_c = 29$ K)
\protect\cite{marsiglio95}.
The curves shown are all for the normal state. Note the ``Drude-like''
peak at the origin at high temperatures, and its absence as the temperature
approaches zero.}
\label{fig4}
\end{figure}

\begin{figure}
\mbox{
\epsfig{figure=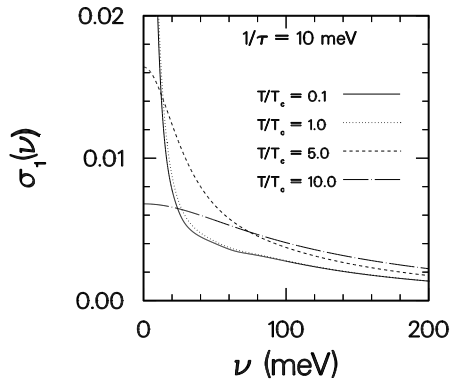,width=0.4\textwidth,clip=}}
\caption{The real part of the optical conductivity as in Fig. 5 but now with
additional elastic scattering ($1/\tau = 10$ meV).
A Drude peak is present at the origin for all temperatures. In both this
figure and in Fig. 4 structure is only discernable at the lowest temperatures.}
\label{fig5}
\end{figure}

\noindent electron-phonon effects from the optical conductivity was made by
Allen \cite{allen71}.  He was motivated by infrared measurements on
superconducting Pb \cite{joyce}, and the possibility of inversion in that case.
The result was a qualitative success, and even yielded quantitatively
good results \cite{farnworth}. Our recent work, which we now describe,
builds on the work of Ref. \cite{allen71}.
 
The remainder of this section concerns only the normal state,
in contrast to tunneling, where the superconducting state was a requirement.
This is both an advantage and disadvantage, as the theory we present is
at zero temperature, which may be difficult to achieve in the normal state
for some materials. The required expressions are \cite{allen71,marsiglio98}:
\be
\sigma(\nu) = {\omega^2_P \over 4\pi} {i \over \nu}
\int_0^\nu d\omega {1 \over \nu + i/\tau - \Sigma(\omega)
- \Sigma(\nu - \omega) }
\label{sigma}
\ee
\noindent where
\be
\Sigma(\omega) = \int_0^\infty d\Omega \alpha^2F(\Omega)
\ln | {\Omega - \omega \over \Omega + \omega} | - i\pi
\int_0^{|\omega|} d \Omega \ \alpha^2F(\Omega)
\label{selfenergy}
\ee
\noindent is the effective electron self-energy due to the electron-phonon
interaction. The spectral function that appears in Eq. (\ref{selfenergy})
is really a closely related function, as has been discussed by Allen \cite{allen71}
and Scher \cite{scher}. For our purposes we will treat them identically.
Eqs. (\ref{sigma}) and (\ref{selfenergy}) elucidate why the electron-phonon
signature is rather modest in the optical conductivity (it is even better
hidden in the reflectance); two integrations effectively average over
the electron-phonon spectral function. One would like to ``unravel'' this
information as much as possible before attempting an inversion, so that,
in effect, the signal is ``enhanced''. To this end one can attempt
various manipulations. As a first step one can make a weak coupling
type of approximation to obtain \cite{marsiglio98} the {\it explicit} result:
\be
\alpha^2F(\nu) = {1 \over 2\pi} {\omega_P^2 \over 4\pi}
{d^2 \over d\nu^2} \biggl\{ \nu Re{1 \over \sigma(\nu)} \biggr\}.
\label{explicit}
\ee
\noindent  Insofar as Eq. (\ref{explicit}) works extremely well (as we shall see
in a moment), it is a remarkable result. It tells us that, with a judicious
manipulation of the conductivity data, the underlying electron-phonon spectral
function emerges in closed form. This is a much better result than we were 
initially hoping for. Not surprisingly it requires two 
derivatives of the data (recall the two integrations), so in practice accurate
measurements are required (actually smoothing the data works quite well ---
and too much smoothing cannot really give you a spurious result; at worst
it will obscure an otherwise noisy result --- see Ref. \cite{puchkov96} for
an early application of Eq. (\ref{explicit})). Perhaps the most inaccurate
part is the determination of the plasma frequency, which can be obtained
through a sum rule or from other measurements, procedures which are both
often fraught with errors. 

Before we evaluate Eq. (\ref{explicit}) with a known example, we note that
a full numerical inversion of Eqs. (\ref{sigma}) and (\ref{selfenergy}) 
is possible. However, a straightforward application of a Newton-Raphson
iteration technique turns out to be very unstable. We derived instead the following
expression \cite{marsiglioerice}:
\be
\alpha^2F(\nu) = {1 \over \pi} \hbox{Im} \,
\Biggl\{
{
2\int_0^{\nu} d\omega \, {
[1 + \lambda(\omega)][1+\lambda(\nu-\omega)]
\over
[\nu + {i \over \tau} - \Sigma(\omega) - \Sigma(\nu - \omega)]^3} +
{4\pi i \over \omega_P^2} {d^2 \over d\nu^2}[\nu \sigma(\nu)]
\over
g^2(\nu) + {1 \over [\nu + {i \over \tau} - \Sigma(\nu)]^2}
}
\Biggr\}
\label{inv1}
\ee
\noindent where
\be
g(\nu) \equiv -{4\pi i \over \omega_P^2} {d \over d\nu}[\nu \sigma(\nu)]
+ \int_0^{\nu} d\omega \,
{ 1 + \lambda(\omega)
\over
[\nu + {i \over \tau} - \Sigma(\omega) - \Sigma(\nu - \omega)]^2},
\label{inv2}
\ee
\noindent and
\be \lambda(\omega) \equiv \int_0^\infty d\Omega
\alpha^2F(\Omega) {2 \Omega \over \Omega^2 - \omega^2}
\label{lambda}
\ee
so that the right hand side depends (in many places through both 
$\lambda(\omega)$ and
$\Sigma(\omega)$) on $\alpha^2F(\Omega)$. 
We have found that only a few iterations of these equations are needed, 
even with an initial guess
which is blank. The result from this procedure in the case of Pb is
plotted in Fig. 6

\begin{figure}
\mbox{
\epsfig{figure=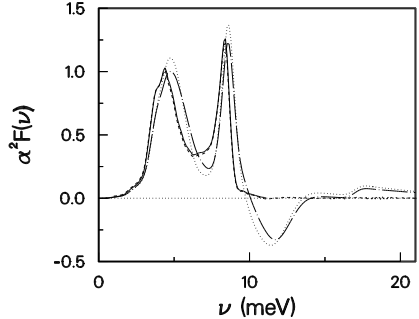,width=0.4\textwidth,clip=}}
\caption{$\alpha^2F(\nu)$ for Pb (solid curve) vs.
$\nu$, along with the estimates obtained from Eq. (\protect\ref{explicit})
with an impurity scattering rate, $1/\tau = 1$ meV (dotted) and
10 meV (dot-dashed).
These are both qualitatively quite accurate, before they
become negative at higher frequencies. Also plotted is the result
(dashed curve, indiscernible from the solid curve)
obtained from a full numerical inversion, as described in the text.}
\label{fig6}
\end{figure}

\noindent (dashed curve), though the curve is 
essentially indistinguishable from the tunneling density of states (solid curve)
which was used to generate the optical conductivity data in the first place.
This illustrates the success of the method. However, the more remarkable
result is that of Eq. (\ref{explicit}). We have also plotted the result of
this formula applied to the conductivity of Pb (again generated numerically)
for two values of the elastic scattering rate (after all in a real experiment
we don't know what this value is). The resulting curves (dotted and dot-dashed)
give very good qualitative results (and don't depend very much on the value
of $1/\tau$ used). They both exhibit a spurious negative part plus some
spurious ``recovery'' noise, so these effects teach us about the limitations
of Eq. (\ref{explicit}). If we ignore such negative pieces, we get a very good
representation of the spectral function. 

\bigskip
\noindent {\it The Case of K$_3$C$_{60}$}
\medskip
\par

For the case of K$_3$C$_{60}$ we simply use the approximate Eq. (\ref{explicit}),
bearing in mind some of the limitations already discussed.
We have used the reflectance data at 25 K for K$_3$C$_{60}$ taken by
Degiorgi {\it et al.} \cite{degiorgi}. Details of our analysis of this
data are given in \cite{marsiglio98}. The result of using Eq. (\ref{explicit})
is plotted in Fig. 7 (solid curve), along with the neutron scattering 
data from \cite{pintschovius} (dashed curve). We also include a result obtained
through analysis of photoemission data \cite{kabanov} (dotted curve), where
we have taken the coupling strengths and frequencies of particular phonon
branches analyzed there, and arbitrarily broadened them.

Our results are in very good qualitative
agreement with the neutron scattering data \cite{pintschovius}.
The energy scale is certainly correct, and moreover the peaks line up fairly 
well, suggesting that the coupling is not dependent on energy. Note that
we have omitted some negative pieces, as these are expected to arise
from the use of Eq. (\ref{explicit}) (not the low frequency negative component,
however, for which we have no satisfactory explanation at present).
We also note that the agreement with the photoemission-derived results 
is not particularly good. \par

We can easily `check' this result by computing the reflectance expected from
this spectral function and comparing to the measured reflectance. The
agreement is within experimental error \cite{marsiglio98}.
Thus, we have demonstrated a consistent explanation of the
data (but not necessarily a unique one).
The fitted parameters are $\omega_P
= 6000 $ cm$^{-1}$ and $1/\tau = 95$ meV for the plasma frequency and
impurity scattering rate, respectively.
\par

\begin{figure}
\mbox{
\epsfig{figure=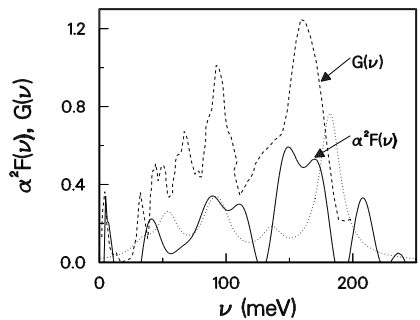,width=0.4\textwidth,clip=}}
\caption{The $\alpha^2F(\nu)$ for K$_3$C$_{60}$ (solid curve)
extracted from the reflectance data of Degiorgi {\it et al.}
\protect\cite{degiorgi},
using Eq. (\protect\ref{explicit}). For purposes of analysis we have omitted
the negative parts. The neutron scattering
results from Ref. \protect\cite{pintschovius} (dashed curve) are also shown.
Clearly the energy
scale in  $\alpha^2F(\nu)$ matches that of the phonons, and some of the
peaks even line up correctly. Finally, the dotted curve comes from an
analysis of photoemission data \protect\cite{kabanov}, where we have
arbitrarily broadened the phonon spectrum with Lorentzian lineshapes.
}
\label{fig7}
\end{figure}

We have used the extracted  $\alpha^2F(\nu)$ in Fig. 2 
to determine whether or not such a spectral function can account
for the superconducting properties in K$_3$C$_{60}$. We find, given 
$T_c=19$ K, the Coulomb repulsion pseudopotential is $\mu^\ast  = 0.4$ 
(using a cut-off, $\omega_c = 1$ eV). This is in effect another check,
since a negative $\mu^\ast$ would indicate a breakdown.
The spectrum is characterized by $\lambda = 1.2$ and $\omega_{ln} =
40$ meV, parameters which give BCS-like results for superconducting
properties (for example, the gap ratio comes out to be
$2\Delta_0/k_BT_c \approx 4$).
If the low frequency peak in $\alpha^2F(\nu)$ at 5 meV is excluded, the
results are then {\it very} BCS-like: $\lambda = 0.8$, $\mu^\ast = 0.34$,
$T_c/\omega_{ln} \approx 0.016$, and the gap ratio is
$2\Delta_0/k_BT_c \approx 3.6$. These results are consistent with some
tunneling and infrared measurements \cite{koller} but inconsistent with
others \cite{zhang}. The degree of coupling roughly agrees with that extracted
through an analysis of photoemission data \cite{kabanov}. 
Microwave measurements may provide a more discriminating probe for the low
frequency part of the spectrum, as discussed in \cite{marsiglio97}. 
\par

In summary, we have described an inversion scheme to extract from the
optical conductivity (or reflectance) a boson
spectral function responsible for inelastic scattering in a metal. 
The full inversion scheme is numerical, but we have also derived an
explicit expression which provides a remarkably good semi-quantitative
result.  We have applied
this technique to K$_3$C$_{60}$, and have found the {\it qualitatively correct}
$\alpha^2F(\nu)$ for this material. Its frequency range lies in the phonon region
and peaks in
$\alpha^2F(\nu)$ line up with peaks in the phonon distribution function as
determined by neutron scattering.  The coupling is sufficiently strong
to explain the superconductivity in this material.
Thus, the weak coupling approach presented here appears to be able to account for
the main features observed in the far-infrared. K$_3$C$_{60}$ appears to be
a conventional weak coupling electron-phonon superconductor.\par

I would like to acknowledge my collaborators in some of this work,
Tatiana Startseva and Jules Carbotte.  This research was supported
by the Natural Sciences and Engineering Research Council (NSERC) of Canada and
by the Canadian Institute for Advanced Research (CIAR).

\end{multicols}

\begin{thebibliography} {999}

\bibitem{hebard}
A.F. Hebard, M.J. Rosseinsky, R.C. Haddon,D.W. Murphy, S.H. Glarum, T.T.M. Palstra,
A.P. Ramirez, and A.R. Kortan, Nature {\bf 350}, 600 (1991).

\bibitem{satpathy}
S. Satpathy, Chem. Phys. Lett. {\bf 130}, 545 (1986).

\bibitem{dresselhaus}
M.S. Dresselhaus, G. Dresselhaus and P.C. Eklund, {\it Science of Fullerenes
and Carbon Nanotubes}, (Academic Press, Toronto, 1996).\par

\bibitem{reviews}
See the reviews and references cited therein:\par
A.P Ramirez, Superconductivity Review {\bf 1}, 1 (1994).\par
M.P. Gelfand, Superconductivity Review {\bf 1}, 103 (1994).\par
W.E. Pickett, Solid State Physics {\bf 48}, 226 (1994).\par
C.H. Pennington and V.A. Stenger, Rev. Mod. Phys. {\bf 68}, 855 (1996).\par
O. Gunnarsson, Rev. Mod. Phys. {\bf 69}, 575 (1997).\par
L. Degiorgi, Advances in Physics, {\bf 47}, 207 (1998).

\bibitem{xiang}
X.D. Xiang, J.G. Hou, G. Briceno, W.A. Vareka, R. Mostovoy, A. Zettl,
V.H. Crespi and M.L. Cohen, Science {\bf 256}, 1190 (1992).\par
X.D. Xiang, J.G. Hou, V.H. Crespi, A. Zettl and M.L. Cohen, Nature {\bf 361},
54 (1993).

\bibitem{zhang}
Z. Zhang, C. Chen and C.M. Lieber, Science {\bf 254}, 1619 (1991).

\bibitem{koller}
D. Koller, M.C. Martin, L. Mih\'aly, G. Mih\'aly, G. Osz\'anyi, G. Baumgartner
and L. Forr\'o, Phys. Rev. Lett. {\bf 77}, 4082 (1996).

\bibitem{degiorgi}
L. Degiorgi, E.J. Nicol, O. Klein, G. Gr\"uner, P. Wachter, S.-M. Huang,
J. Wiley and R.B. Kaner, Phys. Rev. B{\bf 49}, 7012 (1994);
L. Degiorgi et al. Nature {\bf 369}, 541 (1994).

\bibitem{keifl}
R.F. Keifl {\it et al.}, Phys. Rev. Lett. {\bf 70}, 3987 (1993).

\bibitem{uemura}
Y.J. Uemura {\it et al.}, Nature {\bf 352}, 605 (1991).

\bibitem{franck}
J.P. Franck, in {\it Physical Properties of
High Temperature Superconductors IV}, edited by D.M. Ginsberg
(World Scientific, Singapore, 1994) p. 189.

\bibitem{rainer88}
D. Rainer, private communication, although he notes the tunneling
measurement on V with Al as a normal state metal in proximity, from which 
the electron-phonon interaction in Al can be inferred: J. Zasadzinski {\it et al.}
Phys. Rev. B {\bf 25}, 1622 (1982).
 
\bibitem{giaever}
I. Giaever, Phys. Rev. Lett. {\bf 5}, 147 (1960).

\bibitem{swihart}
J.C. Swihart, IBM J. Research Develop. {\bf 6}, 14 (1962).

\bibitem{eliashberg}
G.M. Eliashberg, Zh. Eksperim. i Teor. Fiz. {\bf 38}, 966 (1960);
Soviet Phys. JETP {\bf 11}, 696 (1960).

\bibitem{rowell63}
J.M. Rowell, P.W. Anderson and D.E. Thomas, Phys. Rev. Lett. {\bf 10},
334 (1963).

\bibitem{schrieffer63}
J.R. Schrieffer, D.J. Scalapino  and J.W. Wilkins, Phys. Rev. Lett. {\bf 10},
336 (1963).

\bibitem{mcmillan65}
W.L. McMillan and J.M. Rowell, Phys. Rev. Lett. {\bf 14}, 108 (1965).

\bibitem{mcmillan69}
W.L. McMillan and J.M. Rowell in {\it Superconductivity},
edited by R.D. Parks (Marcel Dekker, New York (1969)) Vol. 1, p.561.

\bibitem{scalapino69}
D.J. Scalapino, in {\it Superconductivity},
edited by R.D. Parks (Marcel Dekker, Inc., New York, 1969)p. 449.

\bibitem {allenmitrovic} P.B. Allen and B. Mitrovi\'{c},
in {\it Solid State Physics,}
edited by H. Ehrenreich, F.~Seitz, and D. Turnbull (Academic, New York,
1982) Vol. 37, p.1.

\bibitem{rainer86}
D. Rainer, in {\it Progress in Low Temperature Physics}, Vol. 10,
edited by D.F. Brewer (North-Holland, 1986), p.371.

\bibitem{carbotte90}
J.P. Carbotte, Rev. Mod. Phys. {\bf 62}, 1027 (1990).

\bibitem{rowell71}
J.M. Rowell and R.C. Dynes, in {\it Phonons, Proceedings of the
international conference, Rennes, France}, edited by M.A. Nusimovici
(Flammarion Sciences, Paris, 1971)p. 150.

\bibitem{ostrick}
J. Ostrick, with R.C. Dynes, has performed some preliminary tunneling measurements
on these systems (private communication).

\bibitem{pintschovius}
L. Pintschovius, Rep. Prog. Phys. {\bf 57}, 473 (1996).

\bibitem{marsiglio98}
F. Marsiglio, T. Startseva and J.P. Carbotte, Phys. Lett. A{\bf 245}, 172 (1998).

\bibitem{timusk}
T. Timusk and D.B. Tanner in {\it Physical Properties of
High Temperature Superconductors I}, edited by D.M. Ginsberg
(World Scientific, Singapore, 1989) p. 339.
D.B. Tanner and T. Timusk in {\it Physical Properties of
High Temperature Superconductors III}, edited by D.M. Ginsberg
(World Scientific, Singapore, 1992) p. 363.

\bibitem{wooten}
F. Wooten, {\it Optical Properties of Solids} (Academic Press, New York,
1972).

\bibitem{marsiglio97}
F. Marsiglio and J.P. Carbotte, Aust. J. Phys. {\bf 50}, 975 (1997);
Aust, J. Phys. {\bf 50}, 1011 (1997).

\bibitem{dolgov91}
O.V. Dolgov, E.G. Maksimov and S.V. Shulga, in {\it Electron-Phonon
Interaction in Oxide Superconductors}, edited by R. Baquero (World
Scientific, Singapore (1991)),p. 30;
S.V. Shulga, O.V. Dolgov and E.G. Maksimov, Physica C {\bf 178}, 266 (1991).

\bibitem{marsiglio95}
F. Marsiglio and J.P. Carbotte, Phys. Rev. B {\bf 52}, 16192 (1995).

\bibitem{dolgov95}
O.V.Dolgov, S.V.Shulga, J. of Superconductivity, {\bf 8}, 611-612 (1995).

\bibitem{marsiglio96}
F. Marsiglio, J.P. Carbotte, A. Puchkov and T. Timusk, Phys. Rev. B{\bf 53}, 
9433 (1996).

\bibitem{holstein54}
T. Holstein, Phys. Rev. {\bf 96}, 535 (1954); Ann. Phys. (N.Y.) {\bf 29},
410 (1964).

\bibitem{allen71}
P.B. Allen, Phys. Rev. B{\bf 3}, 305 (1971).

\bibitem{joyce}
R.R. Joyce and P.L. Richards, Phys. Rev. Letts. {\bf 24}, 1007 (1970).

\bibitem{farnworth}
B. Farnworth and T. Timusk, Phys. Rev. B{\bf 10}, 2799 (1974); ibid
B{\bf 14}, 5119 (1976).

\bibitem{scher}
H. Scher, Phys. Rev. Lett. {\bf 25}, 759 (1970).

\bibitem{puchkov96}
A.V. Puchkov, D.N. Basov and T. Timusk, J. Phys. Condens. Matter {\bf 8},
10049 (1996).

\bibitem{marsiglioerice}
F. Marsiglio, to be published in J. of Superconductivity, Erice conference
proceedings:  Polarons: Condensation, Pairing, Magnetism, June 1998.
 
\bibitem{kabanov}
A.S. Alexandrov and V.V. Kabanov, Phys. Rev. B{\bf 54}, 3655 (1996).

\end{thebibliography}
\end{document}